\documentclass{article}

\usepackage[superscript,biblabel]{cite}
\usepackage[margin=3cm]{geometry}
\usepackage{amssymb,amsfonts,amsmath}
\usepackage{graphicx}
\usepackage{multirow}
\usepackage{color}
\usepackage[normalem]{ulem}
\usepackage{hyperref}

\usepackage[symbol*]{footmisc}
\DefineFNsymbolsTM{myfnsymbols}{
  \textdagger    \dagger
  \textasteriskcentered *
  \textdaggerdbl \ddagger
  \textsection   \mathsection
  \textbardbl    \|%
  \textparagraph \mathparagraph
}%
\setfnsymbol{myfnsymbols}
\newcommand{\abr}[1]{\left\langle#1\right\rangle}
\newcommand{\argmax}{\operatornamewithlimits{argmax}}
\newcommand{\argmin}{\operatornamewithlimits{argmin}}

\newcommand{\eq}[1]{Eq.\ \eqref{eq:#1}}
\newcommand{\eqs}[2]{Eqs.\ \eqref{eq:#1}~and~\eqref{eq:#2}}

\newcommand{\xpeer}{x_{\textrm{peer}}}
\newcommand{\vxpeer}{\vec{x}_{\textrm{peer}}}
\newcommand{\half}{\frac{1}{2}}
\newcommand{\fskew}{f_{skew}}
\newcommand{\flog}{f_{log}}
\newcommand{\xsp}{x^{\star}}
\newcommand{\xsphat}{\hat{x}^{\star}}

\newcommand{\expect}[1]{\mathbb{E}[#1]}

\renewcommand{\figurename}{\textbf{Figure}}
\renewcommand{\tablename}{\textbf{Table}}
\renewcommand{\thefigure}{\textbf{\arabic{figure}}}
\renewcommand{\thetable}{\textbf{\arabic{table}}}

\usepackage{multibib}
\newcites{Methods}{Methods References}
\newcites{SI}{SI References}

\begin{document}

\title{\textbf{The Statistical Mechanics of Human Weight Change}}

\author{John C.~Lang\footnote{Department of Applied Mathematics, University of Waterloo, Canada} 
\and Hans De Sterck\footnote{School of Mathematical Sciences, Monash University, Australia; corresponding author}
\and
Daniel M.~Abrams\footnote{Department of Engineering Sciences and Applied Mathematics, Northwestern University, USA} \footnote{Northwestern Institute for Complex Systems, Northwestern University, USA}}

\date{}
\maketitle 

\noindent
\textbf{In the context of the global obesity epidemic, it is important to know who becomes obese and why.
However, the processes that determine the changing shape of Body Mass Index (BMI) distributions in high-income societies are not well-understood. 
Here we establish the statistical mechanics of human weight change, providing a fundamental new understanding of human weight distributions.
By compiling and analysing the largest data set so far of year-over-year BMI changes, we find, strikingly, that heavy people on average strongly decrease
their weight year-over-year, and light people increase their weight. This drift towards the centre of the BMI distribution is balanced by diffusion resulting from
random fluctuations in diet and physical activity that are, notably, proportional in size to BMI.
We formulate a stochastic mathematical model for BMI dynamics, deriving a theoretical shape for the BMI distribution 
and offering a mechanism to explain the ongoing right-skewed broadening of BMI distributions over time.
The model also provides new quantitative support
for the hypothesis that peer-to-peer social influence plays a measurable role in BMI dynamics.
More broadly, our results demonstrate a remarkable analogy with drift-diffusion mechanisms that are well-known from the physical sciences and finance.}

Over the past 35 years there has been a near doubling in the worldwide prevalence of obesity \cite{WHO2014, FinucaneEtAl2011}.
Obesity is a risk factor for many chronic illnesses \cite{FieldEtAl2001,deGonzalezEtAl2010,AndreyevaEtAl2004}, and has become one of the major public health concerns of our time \cite{WHO2014,FinucaneEtAl2011}. Understanding who becomes obese and why has direct implications in the quest for adequate public health interventions, for example, to determine whether high-risk individuals or the whole population should be targeted
\cite{krishna2015trends,razak2015reply}.
The body mass index (BMI), defined as the mass (in kilograms) divided by the height (in meters) squared, is a standard measure of relative body weight used to classify individuals as underweight ($\mbox{BMI}\leq18.5$), normal weight ($18.5<\mbox{BMI}\leq25$), overweight ($25<\mbox{BMI}\leq30$), or obese ($\mbox{BMI} > 30$). The distribution of BMIs in high-income societies is right-skewed (i.e., skewed towards the high-BMI side) and the mean and standard deviation (SD) have steadily increased over time \cite{PenmanJohnson2006,FlegalTroiano2000,krishna2015trends}, but the causes of the right-skewness and broadening in time are debated \cite{SwinburnEgger2004, PenmanJohnson2006,krishna2015trends,razak2015reply}.
Recent results show that right-skewed broadening of the distribution is not driven by socioeconomic and demographic factors since it occurs equally within social and demographic subgroups
\cite{krishna2015trends}. Therefore, alternative explanations for the broadening are put forward that include variations in genetic susceptibility to obesogenic environmental factors \cite{rosenquist2015cohort,krishna2015trends}, and the ``runaway train'' theory that BMI distributions are right-skewed because high-BMI individuals become subject to a vicious self-reinforcing cycle of weight gain
\cite{SwinburnEgger2004,razak2015reply}.
Also, uncertainty remains over the importance of external factors such as  microbial \cite{TurnbaughEtAl2006} or peer influences 
\cite{ChristakisFowler2007,TrogdonEtAl2008b,Cohen-ColeFletcher2008,PoncelaCasanovasEtAl2015}.

\begin{figure}[!ht]
	\includegraphics[width=\linewidth]{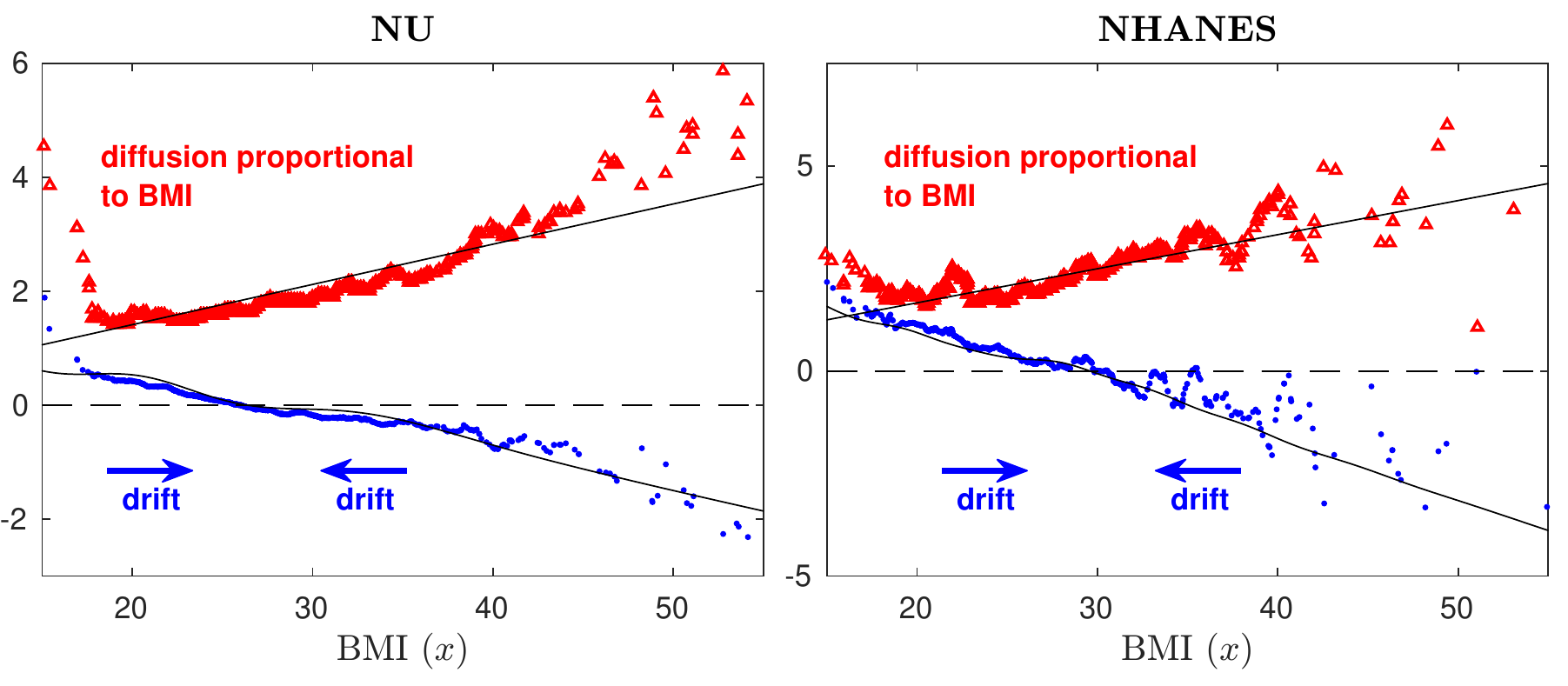}
	\caption{
\textbf{Drift and diffusion in the short-term BMI dynamics of individuals in a human population.}
The figure shows the average annual change in the BMI of individuals (blue dots), and the standard deviation of the annual change in the BMI of individuals (red triangles), as a function of BMI, for data from our new large NU data set (left panel; 121,574 measurements for 2011) and from the publicly available NHANES \cite{NHANES} survey data set (right panel; 5,165 measurements for 2011-2012). The plots are obtained by binning empirical BMI differences (see Methods). The blue curves (dots) show that low-BMI individuals on average increase their weight year-over-year, while high-BMI individuals decrease their weight on average, and the dependence on BMI is approximately linear.  The red curves (triangles) show that the standard deviation of annual BMI changes, which results from natural short-term fluctuations in an individual's BMI that may be due to variations in diet or physical activity, increases approximately linearly as a function of BMI.  These results establish that BMI dynamics feature a \textit{drift} towards a set point, and a \textit{diffusion} that is proportional to the size of the BMI. The thick black curves are the curves of best fit to our mathematical models for the drift term (Eq.~(\ref{eq:Mdetdyn}), including social effects) and for the diffusion amplitude (Eq.~(\ref{eq:Mb})).
\label{fig:2011_Ya_Yb}}
\end{figure}

Figure~\ref{fig:2011_Ya_Yb} presents our main finding: on short timescales of about a year, the BMIs of individuals in a human population show a natural drift \textit{on average} towards the centre of the BMI distribution, and show diffusion (resulting from fluctuations due to multifactorial perturbations) with an amplitude that is approximately proportional to the BMI. We demonstrate this for measurements from two independent data sets (see Methods): we have compiled a new data set of anonymized medical records for more than 750,000 Chicago-area patients of the Northwestern Medicine system (NU) (1997-2014), and we compare with the much smaller but publicly available National Health and Nutrition Examination Survey (NHANES) data set \cite{NHANES} (1999-2012).  

Figure~\ref{fig:2011_Ya_Yb} shows the distinctive trend that \textit{on average} low-BMI individuals increase their weight year-over-year, while high-BMI individuals decrease their weight \textit{on average} (blue dots), with the increase/decrease being approximately linear in BMI. This lends quantitative support to the BMI set point hypothesis: the intrinsic dynamics of weight change in healthy adults are thought to follow a ``return to equilibrium'' pattern where individuals tend to fluctuate about a natural equilibrium, or ``set point'' 
\cite{HallEtAl2011,SpeakmanEtAl2011,KeeseyHirvonen1997}. 
The red triangles in Fig.~\ref{fig:2011_Ya_Yb} show, in a striking manner, that the SD of annual BMI changes increases approximately linearly with BMI. The variation in annual BMI change results from natural short-term fluctuations that may be due to variations in, e.g., diet and physical activity.
The observed linear relation is plausible: high-BMI individuals are expected to lose or gain more weight when subjected to perturbations, e.g., a diet \cite{HallEtAl2011}, for biological reasons \cite{PenmanJohnson2006,razak2015reply}. 

While high-BMI individuals, perhaps surprisingly, decrease their weight \textit{on average}, they are subject to BMI fluctuations with an amplitude (the SD) that is greater than the average decrease in their BMI (Fig.~\ref{fig:2011_Ya_Yb}). The drift towards the centre of the BMI distribution is balanced by these fluctuations, and the fluctuations broaden the distribution away from the centre. This can be understood in analogy with well-known processes from the physical sciences. For example, a massive Brownian particle under the influence of friction due to collisions with molecules in the surrounding medium \cite{Gardiner2004} follows a deterministic path, but at the scale of large populations the collisions between molecules and Brownian particles can be modelled as random fluctuations. The velocity distribution of the Brownian particles can be described accurately by a balance between deterministic drift towards zero velocity (due to friction) and a stochastic diffusion process that models random noise (as described by the Ornstein-Uhlenbeck process), demonstrating that the velocity is normally distributed at equilibrium \cite{Gardiner2004}.
In a similar manner our observations from Fig.~\ref{fig:2011_Ya_Yb} imply that the BMI distribution is intrinsically dynamic and is the result of a balance between deterministic drift and random diffusion, unlike, e.g., the adult height distribution in a human population, which is essentially static on timescales of about a year (because adult height hardly changes). 
We now proceed to describe this balance quantitatively using a stochastic mathematical model.

We model the temporal evolution of the BMI $x_i$ of an individual $i$ by the  Langevin equation \cite{Gardiner2004}
\begin{equation}  
	\label{eq:MLangevin}
	\frac{dx_i}{dt} = a(x_i) + b(x_i) \, \eta(t) \ ,
\end{equation}
where $t$ is time, $a(x_i)$ is a drift (or advection) term and $b(x_i) \, \eta(t)$ forms a random diffusion term
($\eta(t)$ represents Gaussian white noise). 
Since the mean of $dx_i$ is given by $\expect{dx_i} = a(x_i)dt$ and the variance of $dx_i$ by $\expect{dx_i^2} - \expect{dx_i}^2 = b(x_i)^2 dt$, the average of changes in the individual's BMI per time interval $dt$ follows the drift term $a(x)$, and the SD of BMI changes follows $b(x)$. We model the drift term by
\begin{align}
	\label{eq:Mdetdyn}
	a(x_i) &= k_I \, (\xsp-x_i) + k_S \, G(x_i,\vec{x};\sigma).
\end{align}
The first term in Eq.~\eqref{eq:Mdetdyn} represents intrinsic set point dynamics, describing the theory that individuals tend to fluctuate about a natural equilibrium $\xsp$ \cite{HallEtAl2011,SpeakmanEtAl2011,KeeseyHirvonen1997}. 
Our observations of mean annual BMI change in Fig.~\ref{fig:2011_Ya_Yb} suggest a linear relationship with slope $k_I$ as a suitable initial approximation.  The second term of $a(x_i)$ models the extrinsic social influence that individuals may exert on each other, and we base it on the homophily-motivated assumption that individuals interact most strongly with others that are similar 
\cite{Centola2011,blanchflower2009imitative} 
(see Methods). 
We incorporate this effect because our large new data set allows us to investigate whether there is a measurable influence from peer-to-peer dynamics \cite{ChristakisFowler2007,TrogdonEtAl2008b,PoncelaCasanovasEtAl2015}. In the second term, $k_S$ is a rate constant and $G(x_i,\vec{x};\sigma)$ is derived from Gaussian interaction kernels with SD $\sigma$ that model the influence between individual $i$ and the other individuals represented by $\vec{x}$ (see Methods).
Consistent with our observations from Fig.~\ref{fig:2011_Ya_Yb} that fluctuations in an individual's BMI are roughly proportional to BMI, we take 
\begin{equation}
	\label{eq:Mb}
	b(x_i) = \sqrt{k_b} \ x_i\ ,
\end{equation}
with constant $k_b > 0$. In the limit of large population size the aggregate dynamics of individuals are given by a population-level Fokker-Planck equation for the probability density function of the BMI $x$ at time $t$ (see Methods). Since BMI distributions vary slowly on timescales of about a decade, it can be assumed that parameter values in our model drift on a time scale slower than individual equilibration times, and we can therefore consider equilibrium distributions that reflect a balance between the drift and diffusion processes (see Methods).
We thus obtain a closed-form solution for the theoretical BMI distribution without social effects ($k_s=0$ in Eq.~(\ref{eq:Mdetdyn})):
\begin{equation}	
	\label{eq:Mpeq0}
	p_{eq}^{(0)}(x) = c \, x^{-2(k_I/k_b+1)} \, \exp\left(-2\frac{k_I}{k_b}\frac{\xsp}{x}\right),
\end{equation}
where $c$ is a normalization constant (see Methods).
When social effects are included ($k_s\ne0$ in Eq.~(\ref{eq:Mdetdyn})), no closed-form solution exists and the equilibrium distribution must be calculated numerically (see Section~\ref{A:peqIter} in the online Supplementary Information (SI)).

\begin{figure}[!ht]
	\includegraphics[width = 0.45\linewidth]{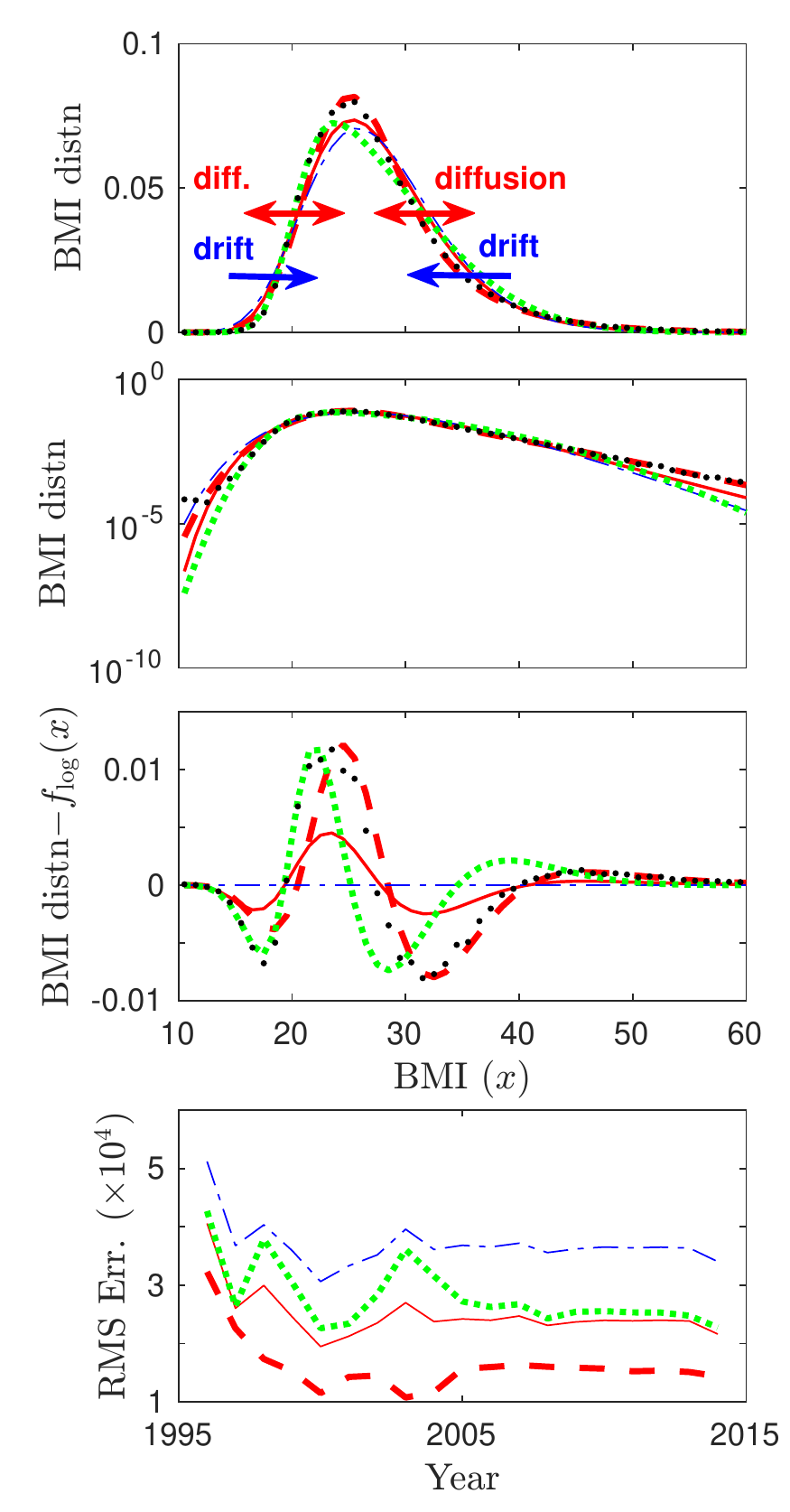}
	\caption{\textbf{
	Results from fitting the 2011 NU empirical BMI distribution (black dots) to predicted distributions $p_{eq}^{(0)}(x)$ (no social effects; red solid) and $p_{eq}(x)$ (with social effects; red dashed), and to standard log-normal (blue dash-dotted) and skew-normal (green dotted) distributions.} 
From top to bottom, the first panel illustrates that the BMI distribution results from a balance between drift and diffusion, and is right-skewed. The second panel shows the same BMI distributions in log scale to make tails more visible, and the third panel shows differences between the log-normal distribution as null-model \cite{PenmanJohnson2006} and the other distributions. The second and third panels show that the $p_{eq}^{(0)}(x)$ (red solid) and $p_{eq}(x)$ (red dashed) distributions are more successful in fitting the empirical data than the commonly used log-normal and skew-normal distributions, both near the centre of the distribution and in the high-BMI tail. This is confirmed in the bottom panel that shows the root mean-square error (RMSE) resulting from fitting NU data to BMI distributions in the range 1997-2014. 
	}
	\label{fig:2011_distn}
\end{figure}

Figure~\ref{fig:2011_distn} shows that our non-social model (two parameters) gives a better fit to empirical BMI distributions than the log-normal (two parameters) and skew-normal (three parameters) distributions that are commonly used to describe right-skewed data. 
However, our social model (four parameters) has the best fit.
These findings are confirmed for publicly available data from the NHANES \cite{NHANES} and BRFSS \cite{BRFSS} surveys in Extended Data Figure~\ref{fig:NHANES-BRFSS-dist} (see also Methods).
To investigate the importance of the social utility contribution to $a(x)$ in \eq{Mdetdyn} we 
compute the relative likelihood ratios of all BMI distribution models 
using the Akaike Information Criterion (AIC) \cite{akaike1974new} (see SI Section \ref{sec:AIC-app}), which quantifies the trade-off between goodness-of-fit and model complexity (number of parameters).
Extended Data Table \ref{tab:AIC_2011}
indicates that our social model is a much better fit to the data than the nonsocial model when taking into account the number of parameters, especially for our large NU data set.
This provides new support for the hypothesis that peer-to-peer social effects play a measurable role in individual BMI dynamics \cite{ChristakisFowler2007,TrogdonEtAl2008b,PoncelaCasanovasEtAl2015}.

Our findings directly offer a new and compelling mechanism to explain the right-skewness of BMI distributions\cite{SwinburnEgger2004,PenmanJohnson2006,krishna2015trends,razak2015reply}: in essence, random fluctuations broaden the BMI distribution away from the set point, and the broadening is stronger on the high-BMI side because the random variations in BMI are proportional to BMI (Fig.~\ref{fig:2011_Ya_Yb}, red triangles).
When explaining the right-skewness, there is thus, at first instance, no need to invoke singular effects such as the ``runaway train'' mechanism\cite{SwinburnEgger2004}, in which high-BMI individuals become subject to a self-reinforcing cycle of weight gain. In fact, we demonstrate that high-BMI individuals on average \textit{strongly decrease} their weight year-over-year (Fig.~\ref{fig:2011_Ya_Yb}, blue dots). However, they are subject to large-amplitude fluctuations (with both positive and negative signs) that broaden the BMI distribution more on the high-BMI side than the low-BMI side. 
Similarly, increasing fluctuations over time also explain the broadening of BMI distributions over time on the high-BMI side \cite{krishna2015trends,razak2015reply} (see Methods).
These fluctuations represent the \textit{aggregate effect} of natural variations in diet and physical activity, and perturbations that result from factors ranging from biology to psychology to social phenomena \cite{PenmanJohnson2006,Foresight2007,krishna2015trends,razak2015reply}, which may indeed include genetic effects\cite{rosenquist2015cohort,krishna2015trends} and self-reinforcing weight gain such as in the ``runaway train'' \cite{SwinburnEgger2004}.
The ultimate reason for the right-skewness can be traced back to the proportionality of BMI fluctuations to BMI, in the balance between drift and diffusion:
individuals are subject to multifactorial perturbations and,
for biological reasons, high-BMI individuals tend to lose or gain more weight due to these perturbations 
\cite{PenmanJohnson2006,HallEtAl2011,razak2015reply}.

In terms of public health interventions, our results indicate that, as the population BMI average increases over time
\cite{SwinburnSacksRavussin2009,MokdadEtAl1999}, the whole population is sensitive to increasing BMI fluctuations (Fig.~\ref{fig:2011_Ya_Yb}, red triangles). These fluctuations ultimately broaden the distribution (especially on the high-BMI side) and increase the high-BMI segment of the population. This adds justification to interventions that target the whole population. On the other hand, we demonstrate and quantify that high-BMI individuals are particularly at risk for large fluctuations that may result from multifactorial perturbations (Fig.~\ref{fig:2011_Ya_Yb}, red triangles), and our results confirm that reducing these fluctuations by discouraging perturbations such as yo-yo dieting \cite{brownell1994medical} should be another focus of intervention.
More broadly, our results establish a form of statistical mechanics for human weight change. Analogous to drift-diffusion processes in physics and finance \cite{Gardiner2004,Shreve2004}, our empirical findings and model provide a new fundamental understanding of the role of drift and diffusion mechanisms in the dynamics of BMI distributions in human populations.

\clearpage

\clearpage

\bibliographystyle{naturemag}
\bibliography{biblio3}

\paragraph{Acknowledgments}
  The authors thank Bonnie Spring and Michael Hynes for useful conversations.  DMA thanks the James S. McDonnell Foundation for their support through grant \#220020230, Anna Pawlowski for her help with data compilation, and the Northwestern Medicine Enterprise Data Warehouse Pilot Data Program for their support (Task ID \#21051).

\paragraph{Author Contributions}
All authors contributed equally to the design of this study and to the writing of the paper. JCL performed the data analysis, simulations and calculations for this work. 

\clearpage
\section*{Methods}

\setcounter{equation}{0}
\renewcommand{\theequation}{M\arabic{equation}}

\paragraph{BMI Data Sets}
For this work we require two different types of BMI data: population-level and individual-level. At the population level we consider empirical BMI distributions over a population (see Fig.~\ref{fig:2011_distn} and Extended Data Figure~\ref{fig:NHANES-BRFSS-dist}). We compute empirical BMI distributions from three independently collected data sets: our new data set of medical records for Chicago-area patients of the Northwestern Medicine system of hospitals and clinics (NU) that we make freely available, and two publicly available data sets that derive from national health surveys in the United States, the National Health and Nutrition Examination Survey (NHANES) \cite{NHANES}, and the Behavioural Risk Factor Surveillance System (BRFSS) \cite{BRFSS}. At the level of individuals we consider the average change in individuals' BMIs over time and the standard deviation in the changes in individuals' BMIs, both as a function of BMI (see Fig.~\ref{fig:2011_Ya_Yb}). We can compute the temporal change in individuals' BMIs from two independently collected data sets: the new NU and the existing NHANES data sets. Our study and model focus on BMI changes of individuals over short timescales, and in practice a suitable timescale for which data on BMI change is available is of the order of about a year, since multiple measurements typically exist for patients visiting hospitals on the time scale of a year, and health survey data also often provide information on annual changes.

\paragraph{New Data Set: Northwestern Medicine Medical Records}
As part of this study, we compile and present analysis of an entirely new BMI data set more abundant than any previously reported. BMI measurements calculated from anonymized medical records for more than 750,000 patients of the Northwestern Medicine system of hospitals and clinics are considered from 1997 through 2014, with the majority of records coming from later years. We calculate BMI from weight and height data for individuals in this data set that are at least 18 years of age. We use these data to compute the empirical BMI distribution for each year. In addition, we are able to calculate the change in BMI over one year for all individuals with patient records in consecutive years. Specifically, we extract from the Northwestern Medicine medical record 1,017,518 measurements of year-over-year BMI change for 329,543 distinct individuals. 
We note that this data set provides the most abundant source of individual level data. 
However, one caveat is that these data do not form a fully representative sample of the population (since they are comprised of medical records). For this reason, we carefully vet all our results and findings by cross-comparison with the NHANES and BRFSS survey data, which can be assumed to be representative of the US population. Nevertheless, our new NU data are extremely valuable since they were recorded during actual physical exams (unlike some of the survey interview data which were self-reported). They represent the largest data set of its type and allow us to conduct more detailed studies. For additional details on the NU data, see Section~\ref{A:NU} in the online Supplementary Information (SI).

\paragraph{Publicly Available NHANES and BRFSS Survey Data}
In SI Sections~\ref{A:NHANES} and \ref{A:BRFSS} we describe the publicly available NHANES and BRFSS survey data. NHANES data are available for survey years 1999-2000, 2001-2002, \ldots, 2011-2012, and allow us to consider empirical BMI distributions based on approximately 5,000 adult individuals per year whose weight and height measurements were taken during a physical exam. 
The NHANES data also provide self-reported change in BMI over the year preceding the survey interview. We consider BRFSS data for survey years from 1987 to 2013. The number of individual records increases from approximately 50,000 in 1987, to more than 400,000 from 2007 onward. Weight and height measurements are self-reported. We use BRFSS data as a third source for empirical BMI distributions, but the BRFSS data does not contain information that allows us to infer annual BMI change for individuals.

\paragraph{Methods for Figure \ref{fig:2011_Ya_Yb}}
The blue dots give the average annual change in the BMI of individuals, as a function of BMI. The averages are taken over bins of empirical BMI differences; BMI differences that originate from a similar starting BMI are placed in the same bin. 
Specifically, to generate Fig.~\ref{fig:2011_Ya_Yb} we first compute average and standard deviation of year-over-year BMI differences on the 90-point grid $\{10.5,11.5,12.5,\ldots,99.5\}$. For each grid point the average and standard deviation of year-over year BMI differences are taken over the bin containing all BMI differences with initial BMI within $\epsilon = \half$ of the grid point. We then compute the average and standard deviation of year-over-year BMI change for individual $i$ by using linear interpolation, i.e. we take a weighted average of the average and standard deviation year-over-year BMI change at the two grid points closest to individual $i$'s initial BMI.
We plot the averages (blue dots) and standard deviations (red triangles) for 1,000 individuals selected uniformly at random from the sampled population.
(We display the values for only 1,000 individuals to improve visualization of data.)
For the 2011 NU data there are 121,574 individual BMI difference measurements and each bin (associated with a point in the grid $\{10.5,11.5,\ldots,99.5\}$) contains on average 1,350 BMI differences. For 2011-2012 NHANES data there are 5,165 individual BMI difference measurements and each bin contains on average 57 BMI differences.
In SI Section \ref{A:ind_params} we explain how we fit the parameters in $a(x)$ (\eq{Mdetdyn}) and $b(x)$ (\eq{Mb}) to obtain the fitted black solid curves in Fig.~\ref{fig:2011_Ya_Yb}. The fitted parameters are presented in Extended Data Table \ref{tab:fitaMain}.

\paragraph{Modeling Intrinsic Drift Dynamics}

The first term in Eq.~\eqref{eq:Mdetdyn} models intrinsic drift dynamics.
The intrinsic dynamics of return to a set point weight is modeled by assuming exponential decay to equilibrium as
\begin{equation}  
	\label{eq:expdecay}
 	\frac{dx_i}{dt} = k_1 (\xsp_i-x_i)~,
\end{equation}
where $\xsp_i$ represents the individual's BMI set point, and the constant $k_1>0$ determines the rate of exponential relaxation to equilibrium weight (note that we assume constant height in adults over time, so changes in BMI---defined as the ratio of weight to height squared---are proportional to weight changes). 
This set point weight may depend upon many factors including genetics, average exercise and eating habits, etc.  Though the set point may vary gradually over the course of an individual's life, we approximate it as a constant on the shorter time scale over which our model applies. In addition, to obtain tractable models, we assume in most of our approach that individuals have a common set point $\xsp$. This is a reasonable first approximation as indicated by the curves of average annual BMI change in Fig.~\ref{fig:2011_Ya_Yb}, which shows that there is a nearly linear variation with an intersection point of the curve that is relatively clearly defined.

Another way to deduce this same model for intrinsic set point dynamics is to assume that individuals tend to maximize some \emph{individual utility function} $u_I(x) = u_I(x;\xsp)$, which by assumption must have a local maximum when BMI $x=\xsp$ and can be modeled in first approximation by a quadratic as in
\begin{equation}
	\label{eq:Taylor_ui}
  	u_I(x) \approx - \frac{1}{2} k_2 (x-\xsp)^2.
\end{equation}
Assuming that the rate of change of BMI will be proportional to the rate of increase of utility,
\begin{equation} 
	\label{eq:intdyn}
  	\frac{dx}{dt} = k'_2 \frac{d u_I}{dx}~,
\end{equation}
we arrive at the same intrinsic dynamics as model \eqref{eq:expdecay} (equations \eqref{eq:expdecay} and \eqref{eq:intdyn} are identical when $k_1= - k'_2 \, k_2$).

\paragraph{Modeling Extrinsic Drift Dynamics}

The second term in Eq.~\eqref{eq:Mdetdyn} models the extrinsic, peer-to-peer social part of the drift dynamics. 
The extrinsic dynamics of weight change are more difficult to model.  Some theories suggests that individuals can become accustomed to the average BMI of peers under exposure to different peer environments \cite{Centola2011,blanchflower2009imitative}
and, to reduce disparity, may adjust their weights \citeMethods{McPhersonEtAl2001,burke2010overweight}.  We assume that there exists some \emph{social utility function} $u_S(x)=u_S(x;\vxpeer)$ which captures this proposed peer-influence phenomenon: the social utility should peak when an individual reaches a BMI consistent with his or her peer(s), $\vxpeer$, where $\vxpeer$ is a vector containing the BMIs of the peers. Similarly to the intrinsic dynamics, we expect this utility to be well approximated, for the case of a single peer, by a quadratic function (at least locally) and therefore propose
\begin{equation} 
	\label{eq:uSoc1}
  	v(x;\xpeer) \approx -\frac{1}{2} k_3 ( \xpeer - x )^2~,
\end{equation}
where we assume that $k_3 > 0$ is a constant, and where $\xpeer$ is the BMI of some peer who influences the individual under consideration.  When multiple peers simultaneously influence an individual, the net social utility becomes
\begin{equation*}
  	u_S(x_i) = u_S(x_i;\vec{x}) \approx - \frac{1}{2} k_3 \sum_{j=1}^N A_{ij} (x_j - x_i)^2~,
\end{equation*}
where $N$ is the number of individuals in the population, $\vec{x} = (x_1,x_2,\ldots,x_N)^T$, and $A_{ij}$ represents the strength of social influence of individual $j$ on individual $i$. Note that we use $v$ to denote the social influence of a single peer and $u$ for the cumulative effect of multiple peers.%

Combining both the intrinsic and extrinsic aspects of the proposed drift process, we obtain
\begin{equation}  
  	\frac{dx_i}{dt} = \frac{du(x_i)}{dx_i} = a(x_i)
\end{equation}
where
\begin{align}
	\label{eq:detdyn}
	a(x_i) &= k_I (\xsp-x_i) \nonumber \\
	&+ k_S \underbrace{\sum_{j=1}^N \left[ A_{ij}(x_j - x_i) - \half \frac{\partial A_{ij}}{\partial x_i} (x_j-x_i)^2 \right]}_{G(x_i,\vec{x};\sigma)},
\end{align}
and the constants $k_I$ and $k_S$ set the relative importance of individual versus social factors. Note that the summation in \eq{detdyn} corresponds to $G(x_i,\vec{x};\sigma)$ in \eq{Mdetdyn}.

In order to specify $A_{ij}$ we make the homophily-motivated assumption \cite{Centola2011,blanchflower2009imitative} that individuals interact most strongly with others who are most similar to themselves, i.e., individuals with similar BMI interact more strongly than individuals with different BMI 
%
\citeMethods{McPhersonEtAl2001,burke2010overweight}.
Consistent with this assumption, we choose a Gaussian interaction kernel
\begin{equation}
	\label{eq:Aij}
	A_{ij} = \frac{1}{N} \phi_{x_i,\sigma}\left( x_j \right),
\end{equation}
where $N$ is the population size, $\sigma > 0$ is a fixed parameter, and $\phi_{\mu,\sigma}(x)$ is the probability density function of a normal random variable with mean $\mu$ and standard deviation $\sigma$ evaluated at $x$. This has the effect of imposing stronger interaction among more similar individuals.

\paragraph{Fokker-Planck Equation and Equilibrium Distribution}  

In the limit of large population size $N\rightarrow\infty$, the aggregate dynamics of individuals described by Langevin equation (\ref{eq:MLangevin}) are given by the population-level Fokker-Planck equation \cite{Gardiner2004}
\begin{equation}
	\label{eq:2peqSolve}
	\frac{\partial p}{\partial t}(x,t) = -\frac{\partial }{\partial x}[p(x,t)a(x)] + \frac{1}{2}\frac{\partial^2 }{\partial x^2}[p(x,t)b^2(x)],
\end{equation}
where $p(x,t)$ is the probability density function for BMI $x$ at time $t$. The correspondence with the Langevin equation is exact when $k_S=0$ (no social effects), and we assume that it holds in first approximation otherwise since social effects are a relatively small correction to the dominant linear trend of the drift term $a(x)$.
Since BMI distributions vary slowly on timescales of about a decade, it can be assumed that parameter values in our model drift on a time scale slower than individual equilibration times. We can therefore consider equilibrium distributions, and 
setting the time derivative to zero we obtain the closed-form solution of \eq{Mpeq0} for $p_{eq}^{(0)}(x)$, the theoretical BMI distribution without social effects ($k_s=0$ in Eq.~(\ref{eq:Mdetdyn})).
When social effects are included ($k_s\ne0$ in Eq.~(\ref{eq:Mdetdyn})), no closed-form solution exists and the equilibrium distribution must be calculated numerically (see Section~\ref{A:peqIter} in the online Supplementary Information (SI)).
The normalisation constant in the closed-form formula for the non-social theoretical BMI distribution,
$p_{eq}^{(0)}(x)$ (\eq{Mpeq0}), is given by
$$ c^{-1} = \left( 2 \xsp \frac{k_I}{k_b} \right)^{-2k_I/k_b-1} \Gamma(2 k_I/k_b + 1),$$
where $\Gamma(t) = \int_0^{\infty} x^{t-1}e^x dx$ is the Gamma function.

We note that since $p_{eq}^{(0)}(x) \sim x^{-2(k_I/k_b+1)}$ as $x\rightarrow\infty$, $p_{eq}^{(0)}(x)$ becomes a scale-free (or power law) distribution.
Note that the linear assumption of \eq{Mb} also naturally implies a vital property of the equilibrium distribution in our model, namely, that the probability is confined to positive BMIs. Indeed, diffusion of probability is halted at $x=0$.
In order for the variance to be non-negative we require that $2 k_I/k_b - 1 > 0$. This condition is satisfied in all the empirical BMI distributions considered in this study.
The mean, mode, variance, skewness, and mode skewness of this distribution can be expressed in terms of $\xsp$ and $k_0 = k_I/k_b$ and are recorded in Extended Data Table~\ref{tab:p0prop}, see SI Section~\ref{A:p0prop} for detailed calculations.

\paragraph{Methods for Figure \ref{fig:2011_distn}}  

In Fig.~\ref{fig:2011_distn} and Extended Data Figure~\ref{fig:NHANES-BRFSS-dist} we compare our new theoretical BMI distributions with two other candidate distribution functions that are commonly used to describe right-skewed data: the log-normal probability distribution function
\begin{equation}
	\label{eq:flog}
	\flog(x;\mu,\sigma) = \frac{1}{x\sqrt{2\pi\sigma^2}} \exp\left[-\frac{(\log x -\mu)^2}{2\sigma^2}\right],
\end{equation}
and the skew-normal probability distribution function 
\begin{equation}
	\label{eq:fskew}
	\fskew(x;\xi,\omega,\alpha) = \frac{2}{\omega} \phi_{0,1}\left( \frac{x-\xi}{\omega} \right) \Phi_{0,1} \left[ \alpha\left(\frac{x-\xi}{\omega}\right) \right],
\end{equation}
where $\Phi_{\mu,\sigma}(\cdot)$ is the cumulative distribution function for a normal random variable with mean $\mu$ and standard deviation $\sigma$. For details on how we fit empirical BMI distributions, see SI Section \ref{A:dist_fit}.

\paragraph{Right-Skewness and Broadening of the BMI Distribution over Time}  

Our results offer a mechanism to explain why BMI distributions continue to broaden over time, especially on the high-BMI side \cite{krishna2015trends,razak2015reply}. Essentially, in the context of our findings the observed growth in average BMI (Extended Data Fig.~\ref{fig:empirical_BRFSS}) implies more fluctuations since fluctuations are proportional to BMI (Fig.~\ref{fig:2011_Ya_Yb}, red triangles), and more fluctuations mean a broadening of the distribution. In addition, we observe a decrease over time of $k_0=k_I/k_b$, which reflects a growing relative importance of fluctuations over drift (see Extended Data Fig.~\ref{fig:k0_BRFSS} and Methods), and we will explain now that this also implies a broadening of the distribution, especially on the high-BMI side.

Extended Data Fig.~\ref{fig:empirical_BRFSS} illustrates that BMI mean and SD have both steadily grown since at least 1987 while the obesity epidemic was running its course (with tempered growth in more recent years) \cite{WHO2014,FinucaneEtAl2011,PenmanJohnson2006,krishna2015trends}.
The expressions for the SD and skewness of our theoretical BMI distribution of \eq{Mpeq0} (see Extended Data Table \ref{tab:p0prop})
serve to further clarify and quantify how our empirical findings and model explain why BMI distributions are right-skewed, and why the right-skewness and SD continue to increase over time.

In terms of explaining why SDs of US BMI distributions continue to increase over time and why BMI distributions broaden \cite{krishna2015trends,razak2015reply},
the formula for the SD in our theoretical BMI distribution of \eq{Mpeq0}, given by ${\xsp}/\sqrt{2k_0-1}$ (Extended Data Table \ref{tab:p0prop}), provides insight. SD increases proportionally with the mean BMI, and a decrease in $k_0$ (increasing importance of fluctuations) also implies an increase in SD.
Intuitively, an increase in the mean BMI implies more fluctuations since fluctuations are proportional to BMI, and a decrease in $k_0$ (the relative importance of drift over fluctuations) also implies more fluctuations. These increasing fluctuations naturally broaden the BMI distribution over time.

The skewness of our theoretical BMI distribution \eq{Mpeq0} is given by $2\sqrt{2k_0-1}/(k_0-1)$ (Extended Data Table \ref{tab:p0prop}). 
Note that positive values for skewness correspond to right-skewness.
Extended Data Fig.~\ref{fig:k0_BRFSS} shows that $k_0=k_I/k_b$, which reflects the relative importance of drift over fluctuations, has steadily decreased over the course of the obesity epidemic, at least since 1987. This decrease is likely due in large part to an increase in the fluctuation proportionality constant $\sqrt{k_b}$ over time, which may plausibly be linked to the factors that have caused the increase in average BMI for the population over time, for example, an increase in average calorie intake or portion sizes over time \citeMethods{young2002contribution}. Indeed, if human BMIs are characterised by short-term fluctuations (Fig.~\ref{fig:2011_Ya_Yb}, red triangles), one can expect these fluctuations to become larger when average calorie intake or portion sizes increase over time.
Applying skewness formula $2\sqrt{2k_0-1}/(k_0-1)$ to the fitted values of $k_0$ in Extended Data Fig.~\ref{fig:k0_BRFSS}, one finds, for example,
that the skewness $\approx 0.77$ for $k_0\approx15$ (for 1996), and skewness increases as $k_0$ decreases over time. This shows that our predicted BMI distribution naturally features right-skewness (essentially due to the fluctuations being larger on the high-BMI side), and that skewness increases over time (since $k_0$ decreases). 

To increase our understanding of the obesity epidemic, it is crucially important to further investigate the detailed influence of factors like genetic susceptibility \cite{rosenquist2015cohort}, and to study who in the BMI distribution is gaining weight \cite{razak2015reply}.
Nevertheless, when considering the \textit{aggregate} effect of factors that influence weight gain, our analysis demonstrates that fluctuations result that are, remarkably, approximately proportional to BMI. Our results show that this proportionality in itself, in the context of the drift-diffusion dynamics we have discovered, is an important determinant in explaining why BMI distributions are right-skewed and why they feature right-skewed broadening over time.


\bibliographystyleMethods{naturemag}
\bibliographyMethods{biblio3}

\clearpage
\section*{Extended Data Figures and Tables}

\setcounter{figure}{0}
\renewcommand{\figurename}{\textbf{Extended Data Figure}}
\renewcommand{\thefigure}{\textbf{\arabic{figure}}}

\setcounter{table}{0}
\renewcommand{\tablename}{\textbf{Extended Data Table}}
\renewcommand{\thetable}{\textbf{\arabic{table}}}

\begin{figure}[!ht]
	\centering
	\includegraphics[width = 110mm]{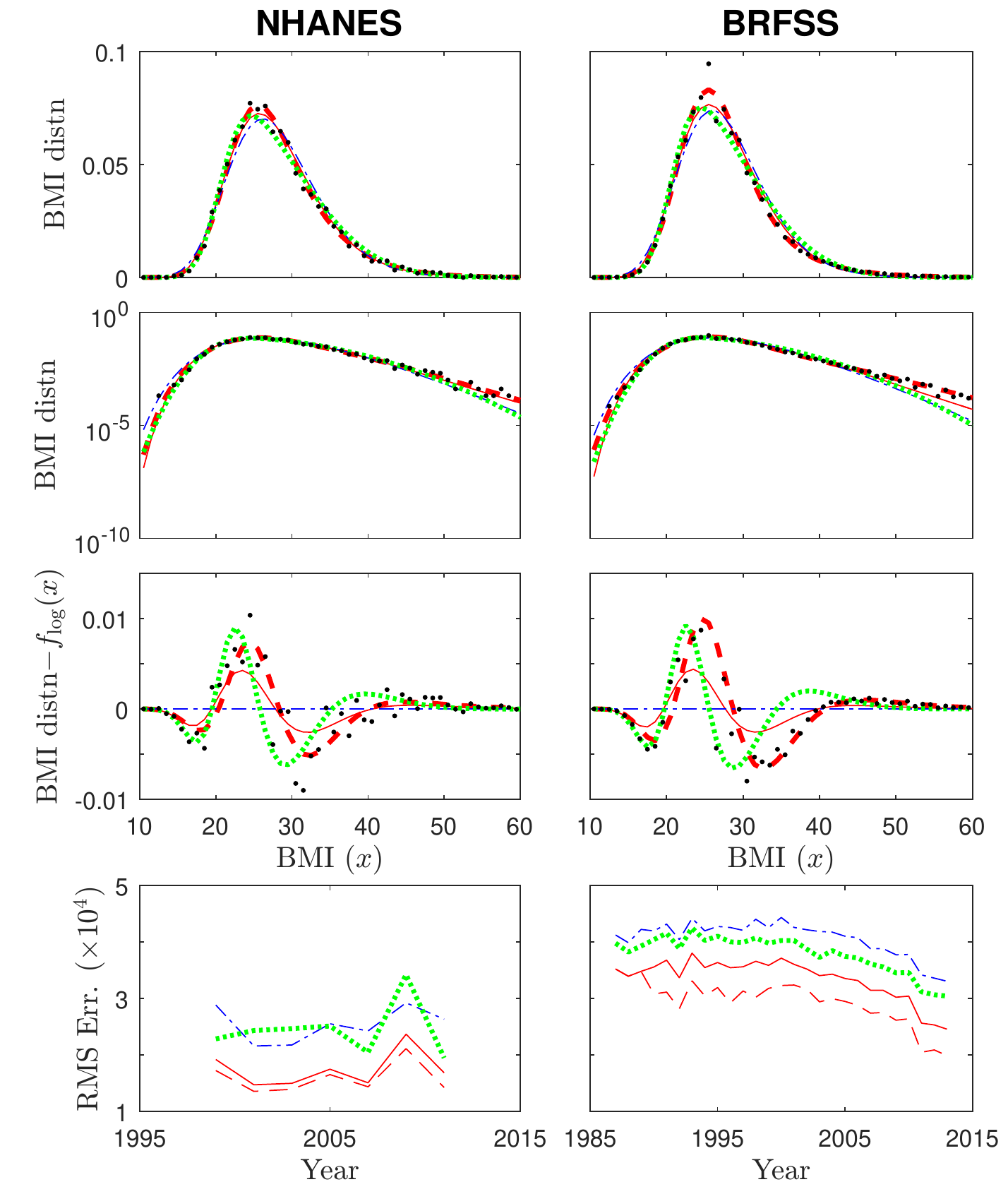} 
	\caption{\textbf{Results from fitting the 2011 NHANES\cite{NHANES} and BRFSS\cite{BRFSS} empirical BMI distributions (black dots) to predicted distributions $p_{eq}^{(0)}(x)$ (no social effects; red solid) and $p_{eq}(x)$ (with social effects; red dashed), and to standard log-normal (blue dash-dotted) and skew-normal (green dotted) distributions.}
The top panels illustrate that the BMI distribution is right-skewed. The second-line panels show the BMI distributions in log scale, and the third-line panels show the difference between the log-normal distribution as null-model\cite{PenmanJohnson2006} and the other distributions. The second-line and third-line panels show clearly that the new $p_{eq}^{(0)}(x)$ and $p_{eq}(x)$ distributions are more successful in fitting the empirical data than the commonly used log-normal and skew-normal distributions. The non-social $p_{eq}^{(0)}(x)$ (red solid), and, in particular the social $p_{eq}(x)$ (red dashed), are a much better fit to the empirical data than the two standard distributions, both in the central part of the distribution (third-line panels) and in the high-BMI tail (second-line panels). Note that the improvement of the social model is less pronounced in the NHANES data, which is likely due to the very small sample size in the NHANES data that appears insufficient to reveal peer-to-peer social effects. This is confirmed in the bottom panels that show the root mean-square error (RMSE) for the data over the full range of years. Overall, the NHANES and BRFSS results are fully consistent with the observations in the main paper for the more extensive NU data.}
	\label{fig:NHANES-BRFSS-dist}
\end{figure}

\begin{figure}[!ht]
	\centering
	\includegraphics[width = 120mm]{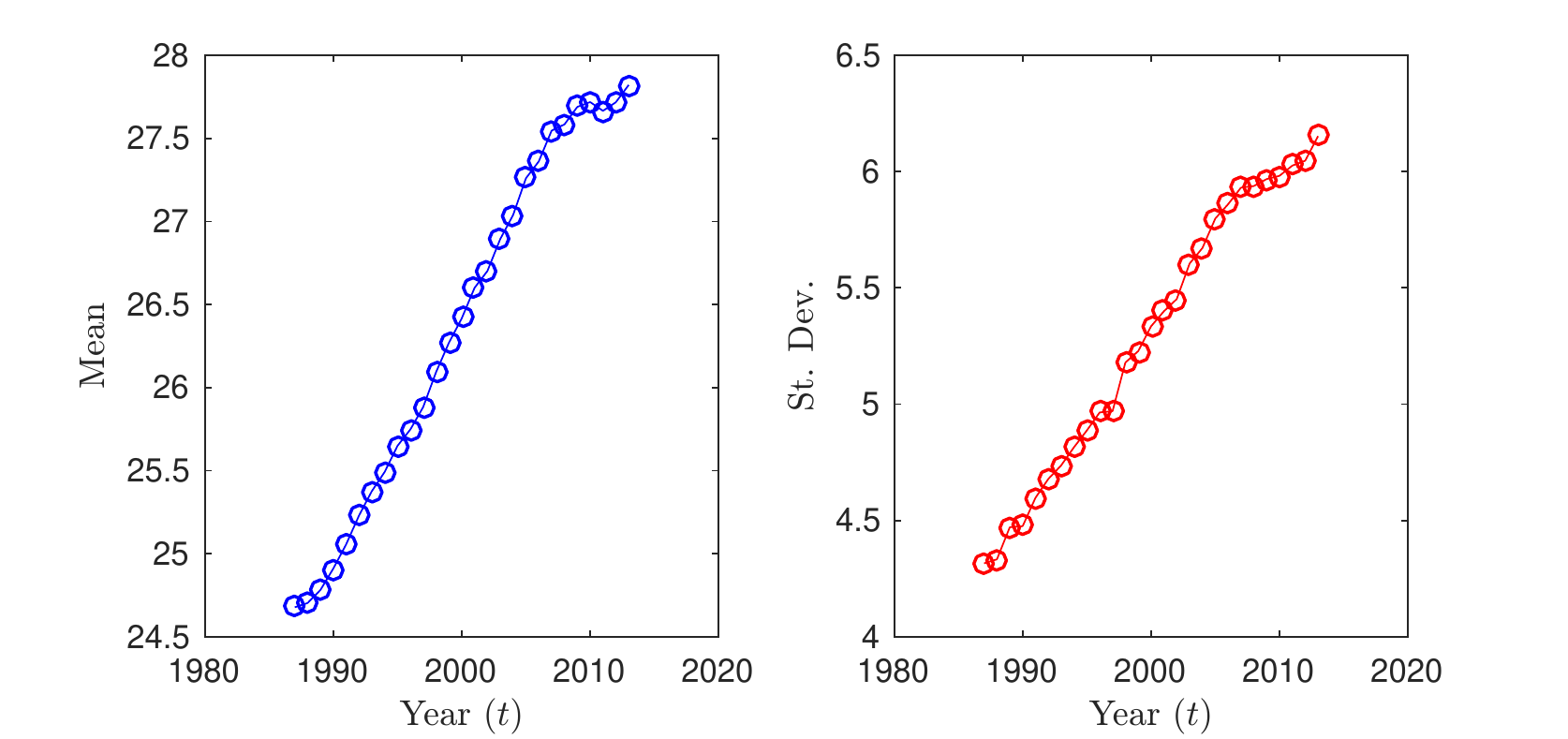} 
	\caption{\textbf{Empirical mean and standard deviation of the BMI for each available year of the BRFSS\cite{BRFSS} survey data.}
	BMI mean and SD have steadily increased over the course of the obesity epidemic, with growth tempered in recent years.}
	\label{fig:empirical_BRFSS}
\end{figure}

\begin{figure}[!ht]
	\centering
	\includegraphics[width = 70mm]{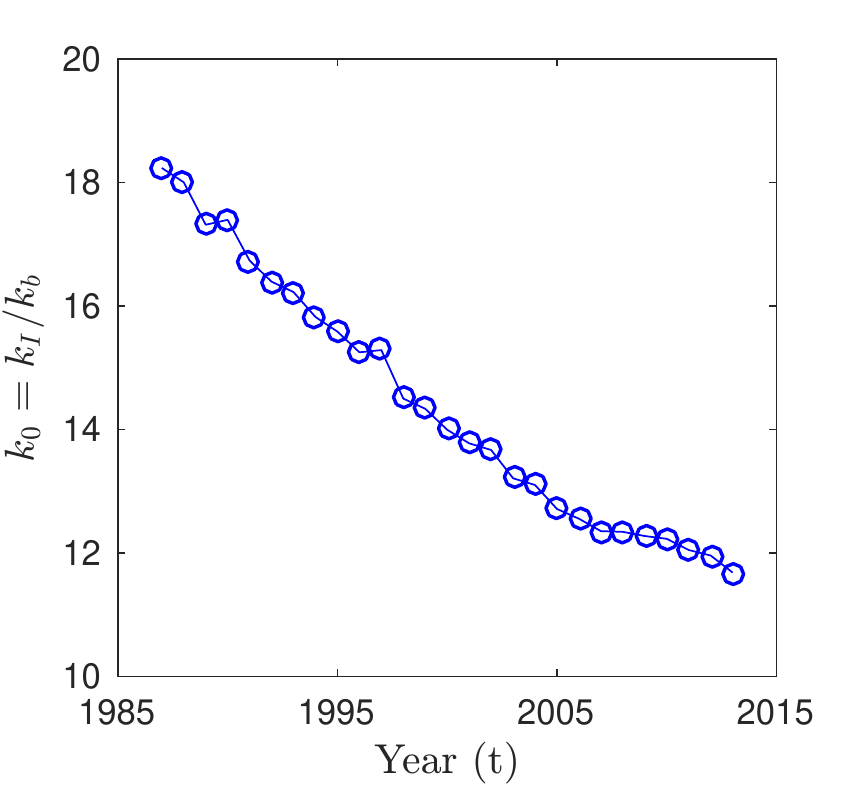} 
	\caption{\textbf{Fitted parameter $k_0=k_I/k_b$ for each available year of the BRFSS\cite{BRFSS} survey data (nonsocial model).} The relative importance of drift over fluctuations has steadily decreased in the course of the obesity epidemic.}
	\label{fig:k0_BRFSS}
\end{figure}

\clearpage

\begin{table}[hp]
	\centering
	\caption{\textbf{Akaike Information Criterion test for model distributions fitted to 2011 empirical BMI distribution data in Fig.~\ref{fig:2011_distn} and Extended Data Fig.~\ref{fig:NHANES-BRFSS-dist}.} Relative likelihood ratio of non-social $p_{eq}^{(0)}(x)$, social $p_{eq}(x)$, log-normal $\flog(x)$, and skew-normal $\fskew(x)$ models for 2011 NU, NHANES\cite{NHANES} and BRFSS\cite{BRFSS} empirical BMI distributions.}
	\begin{tabular}{ccccc}
		\hline
		& \multicolumn{4}{c}{Relative Likelihood Ratio $\exp[(AIC_{\min} - AIC)/2]$}\\
		\cline{2-5}
		Data 		& $p_{eq}^{{0}}(x)$ & $p_{eq}(x)$ & $f_{log}(x)$ & $f_{skew}(x)$  \\ \hline\hline
		NU		& $<10^{-300}$		& 1 	& $<10^{-300}$			& $<10^{-300}$ \\
		NHANES	& $1.3\times10^{-2}$	& 1 	& $1.4\times 10^{-21}$	& $1.1\times 10^{-9}$ \\
		BRFSS	& $<10^{-300}$		& 1 	& $<10^{-300}$			& $<10^{-300}$\\ \hline
	\end{tabular} 
	\label{tab:AIC_2011}
\end{table}

\begin{table}[ht]
	\centering
	\caption{\textbf{Properties of our BMI equilibrium distribution $p^{(0)}_{eq}(x;k_0,\xsp)$, \eq{Mpeq0} ($k_0=k_I/k_b$; no social interaction).}}
	\begin{tabular}{lc}
		\hline
		Property	& Value\\\hline\hline
		Mean 	& $\xsp$\\
		Mode	& $k_0\xsp/(k_0+1)$\\
		Standard Deviation	& ${\xsp}/\sqrt{2k_0-1}$\\
		Skewness			& $2\sqrt{2k_0-1}/(k_0-1)$\\
		Mode Skewness		& $\sqrt{2k_0-1}(k_0+1)$\\ \hline
	\end{tabular}
	\label{tab:p0prop}
\end{table}

\begin{table}[ht]
	\centering
	\caption{\textbf{Parameter estimates for the drift and diffusion curves (solid black) in Fig.~\ref{fig:2011_Ya_Yb}, fitted to the individual-level NU and NHANES\cite{NHANES} data for all available years.} We estimate $k_I$, $\xsp$, $k_S$, and $\sigma$ in the drift term $a(x)$ (\eq{Mdetdyn}) for both the nonsocial ($k_S=0$) and social ($k_S\ne0$) models. We also estimate $\sqrt{k_b}$ in the diffusion amplitude $b(x)$ (\eq{Mb}). The fitting procedure is described in SI Section \ref{A:ind_params}.}
	\begin{tabular}{lccccccccc}
		\hline
		\multirow{2}{*}{Data}	& \multirow{2}{*}{Model} & \multicolumn{4}{c}{Parameters} & \multirow{2}{*}{$L_2$-Error} & \multirow{2}{*}{$\sqrt{k_b}$} & \multirow{2}{*}{$L_2$-Error} \\ \cline{3-6}
						& & $\hat{k}_I$ 			& $\hat{\xsp}$ 		& $\hat{k}_S$ 		& $\hat{\sigma}$		&  && \\ \hline
		\multirow{2}{*}{NHANES}	& nonsocial 	& 0.124		& 28.0	& 0	 	& -- 		& 0.542 & \multirow{2}{*}{$0.083$} & \multirow{2}{*}{0.482}\\
						& social 	& 0.144		& 27.9	& 21.5	& 2.18	& 0.525&&\\\hline
		\multirow{2}{*}{NU}	& nonsocial 	& 0.059		& 28.0	& 0		& -- 		& 0.374 & \multirow{2}{*}{$0.071$} & \multirow{2}{*}{0.461}\\
						& social 	& 0.069		& 28.0	& 9.4 		& 3.44	& 0.306&&\\ \hline	\end{tabular}
	\label{tab:fitaMain}
\end{table}


\clearpage
\section*{Supplementary Information}

\appendix

\setcounter{equation}{0}
\renewcommand{\theequation}{S\arabic{equation}}

This Supplementary Information section contains further information on Data, Methods, and the data and code files that we make available with this manuscript.
\section{Data}

\subsection{Northwestern Medicine Medical Records}
\label{A:NU}

The NU data set consists of medical records from the Northwestern Medical system of hospitals and clinics,  i.e., patients of Northwestern Memorial Hospital, Lake Forest Hospital, and 15 other Chicago area locations: Bucktown (1776 N. Milwaukee Avenue, ​Chicago, Illinois 60647), Deerfield (350 S. Waukegan Road Suites 100, 150 and 200, Deerfield, Illinois 60015), Delano Court (in the Roosevelt Collection, 1135 S. Delano Court Suite A201, Chicago, IL 60605), Evanston (1704 Maple Avenue Suites 100 and 200, Evanston, Illinois 60201), Glenview (2701 Patriot Boulevard, Glenview, Illinois 60026), Grayslake (1475 E. Belvidere Road, Pavilion C Suite 385, Grayslake, IL 60030), Highland Park (600 Central Avenue Suite 333, Highland Park, Illinois 60035), Libertyville (1800 Hollister Drive Suite 102, Libertyville, Illinois 60048), Lakeview (1333 W. Belmont Avenue Suites 100 and 200, Chicago, Illinois 60657), Loop 1 (20 S. Clark Street Suite 1100, Chicago, Illinois 60603), Loop 2 (111 W. Washington St. Suite 1801, Chicago, Illinois 60602), River North (635 N. Dearborn Street Suite 100, Chicago, Illinois 60654), Sauganash (4801 W Peterson, Suite 406, Chicago, IL 60646), Skokie (10024 Skokie Blvd Suite 304, Skokie, IL, 60077), and SoNo (South of North Avenue, 1460 N. Halsted Street Suites 203, 502, and 504, Chicago, Illinois 60642).  

We note that the NU data set may contain multiple measurements per individual per year. In that case the BMI for individual $i$ in year $t$ is calculated using the average weight of individual $i$ in year $t$ and the average height of individual $i$ taken over all years.

For the purpose of computing year-over-year BMI changes, the Northwestern Medicine medical record contains measurements for 329,453 distinct individuals whose BMI can be calculated at at least two time points (1,017,518 BMI differences in total).

\subsection{National Health and Nurtition Examination Survey}
\label{A:NHANES}

The National Health and Nutrition Examination Survey (NHANES) refers to a series of studies designed to collect a representative sample of health and nutrition data for both adults and children (approximately 5,000 individuals total per year) in the United States \cite{NHANES}. NHANES data are available for survey years 1999-2000, 2001-2002, \ldots, 2011-2012. Directly measured BMI data are available from measurements taken during a physical exam. These data are used to compute empirical BMI distributions for each survey year. In addition, during an interview individuals were asked to self-report their current weight and height, as well as their weight from the preceding year. These measurements allow us to calculate self-reported change in BMI over the year preceding the interview. Note: we only use NHANES data for individuals 18 years or older at the time of the survey.

NHANES data are available from the NHANES website
\begin{center} \url{http://www.cdc.gov/nchs/nhanes/nhanes_questionnaires.htm}. \end{center}
Directly measured BMI measurements are given by the variable BMXBMI. Self-reported BMI measurements are calculated from the variables WHD010 (self-reported height at time of interview) and WHD020 (self-reported weight at time of interview). Self-reported change in BMI over the year preceding the interview are calculated from self-reported BMI and from variables WHD010 and WHD050 (self-reported weight one year prior to interview). 

Data were downloaded directly from the Centers for Disease Control  and Prevention (CDC) website as ``.XPT'' files (in SAS format) and imported into Matlab. The variable BMXBMI is found in data files with names starting with ``BMX'', the variables WHD010, WHD020, and WHD050 are found in data files with names starting with ``WHQ'', the variable RIDAGEYR is found in data files with names starting with ``DEMO'', and the SEQN variable is found in all data files. File names are completed by adding the suffix ``.XPT'' for survey year 1999-2000, ``\_B.XPT'' for survey year 2001-2002, ``\_C.XPT'' for survey year 2003-2004, etc...

\subsection{Behavioral Risk Factor Surveillance System}
\label{A:BRFSS}

The Behavioral Risk Factor Surveillance System (BRFSS) refers to a series of telephone surveys designed to collect a representative sample of health data for adults (aged 18 years or older) in the United States \cite{BRFSS}. BRFSS data are available for survey years 1984, 1985, \ldots, 2013. We note that prior to 2011 BRFSS surveys were conducted over land lines only, whereas from 2011 onward BRFSS methodology has been modified to include cell phones as well. We also note that many states did not participate in early BRFSS surveys. Therefore, for the purposes of this study we only consider surveys from 1987 (the first year where a majority of states participated in the BRFSS) onwards. The number of individual records for each BRFSS survey increases from approximately 50,000 in 1987, to approximately 135,000 in 1997, to more than 400,000 from 2007 onward. 
For each BRFSS survey we extract the BMI of each individual surveyed and use this data to compute the empirical BMI distribution for that year. We note that since these data are gathered using telephone interviews, the weight and height measurements (used in calculating BMI) are all self-reported, in contrast to the NHANES and NU data sets. Also in contrast to NHANES and NU data sets, the BRFSS data does not provide sufficient data for us to compute the change in individuals' BMI over time.

BRFSS data are available from the BRFSS website
\begin{center} \url{http://www.cdc.gov/brfss/annual_data/annual_data.htm}. \end{center}
BRFSS surveys record BMI measurements in variable \_BMI for survey years 1984-1999, \_BMI2 for survey years 2000-2002, \_BMI3 for survey year 2003, \_BMI4 for survey years 2004-2010, and \_BMI5 for survey years 2011 onwards. Data were downloaded directly from the CDC website as ``.XPT'' files (in SAS format) and imported into Matlab. File names for BRFSS survey data for years 1978--2010 start with ``CDBRFS'', while file names for BRFSS survey data for years 2011--2013 start with ``LLCP''. File names are completed by adding the suffix ``87.XPT' for year 1987, ``88.XPT'' for year 1988, etc...

\newpage
\section{Additional Details for Methods} \label{A:methodsdetails}
\subsection{Properties of $p^{(0)}_{eq}(x;k_0,\xsp)$ (\eq{Mpeq0})} 
\label{A:p0prop}

The properties of $p^{(0)}_{eq}(x;k_0,\xsp)$ (\eq{Mpeq0}) listed in Extended Data Table \ref{tab:p0prop} can be derived as follows.

We note that for any population the BMI distribution must be strictly contained in the interval $[0,\infty)$. This implies that $p_{eq}(0) = 0$ and that $\lim_{x\rightarrow\infty} p_{eq}(x) =  0$. Assuming that $b(0)=0$ (which holds for our model, see \eq{Mb}), it follows that integrating \eq{2peqSolve} with vanishing temporal derivative yields
\begin{equation}
	\label{eq:peqSolve2}
	0 = -p_{eq}(x)a(x) + \frac{1}{2}\frac{d}{dx}\left[ p_{eq}(x)b^2(x) \right] + \underbrace{ p_{eq}(0)a(0) - \frac{1}{2}\frac{d}{dx}\left[ p_{eq}(x) b^2(x) \right]_{x=0} }_{=0}\ ,
\end{equation}
which has the solution
\begin{equation}
	\label{eq:peqSoln}
	p_{eq}(x) = \xi \exp \left( 2 \int_0^x \frac{a(\tilde{x}) - b(\tilde{x}) b'(\tilde{x})}{b^2(\tilde{x})} d\tilde{x} \right) \ ,
\end{equation}
where $\xi$ is a normalization constant such that $\int_0^\infty p_{eq}(x)dx = 1$. When  $a(x) = k_I(\xsp-x)$ (no social effects, i.e., $k_S=0$ in \eq{Mdetdyn}) and $b(x) =\sqrt{ k_b}\  x$, 
we can re-arrange \eq{peqSolve2} to yield
\[
	\frac{dp_{eq}^{(0)}}{dx}(x) = 2 \frac{k_I(\xsp-x) - k_b x}{k_b x^2} p_{eq}^{(0)}(x) \ ,
\]
which implies that $p_{eq}^{(0)}(x)$ is a single peaked probability distribution whose mode
is given by the expression $\xsp\frac{k_I}{k_b}/(\frac{k_I}{k_b}+1).$ 
(The mode of a continuous random variable with probability density function $f(x)$ is $\argmax_x f(x)$.)
We can also re-arrange \eq{peqSolve2} to yield
\[
	x p_{eq}^{(0)}(x) = -\frac{k_b}{k_I}\frac{d}{dx}\left[\frac{x^2}{2}p_{eq}^{(0)}(x)\right] + \xsp p_{eq}^{(0)}(x)\ ,
\]
which implies that
\[
	\abr{x} = \int_0^{\infty} x p_{eq}^{(0)}(x)dx = -\frac{k_b}{k_I} \underbrace{ \int_0^{\infty} \frac{d}{dx}\left[\frac{1}{2}x^2 p_{eq}^{(0)}(x)\right] dx }_{=0} + \xsp \underbrace{\int_0^{\infty} p_{eq}^{(0)}(x) dx }_{=1}= \xsp \ .
\]
Multiplying \eq{peqSolve2} by $x$ and re-arranging yields
\[
	x^2 p_{eq}^{(0)} (x) = -\frac{1}{2}\frac{k_b}{k_I}\frac{d}{dx}\left[x^3p_{eq}^{(0)}(x)\right] + \frac{1}{2} \frac{k_b}{k_I}x^2 p_{eq}^{(0)}(x) + x \xsp p_{eq}^{(0)}(x) \ ,
\]
which implies
\begin{align*}
	\abr{x^2} &= \int_0^\infty x^2p_{eq}^{(0)}(x)dx \\
	& = -\frac{1}{2}\frac{k_b}{k_I}\int_0^{\infty} \frac{d}{dx}\left[x^3 p_{eq}^{(0)}(x)\right] dx + \frac{1}{2}\frac{k_b}{k_I} \int_0^{\infty} x^2p_{eq}^{(0)}(x)dx + \xsp\int_0^{\infty} xp_{eq}^{(0)}(x)dx \\
	& = \frac{1}{2}\frac{k_b}{k_I} \abr{x^2} + {\xsp}^2\ . 
\end{align*}
Re-arranging now yields $\abr{x^2} = 2{\xsp}^2 \frac{k_I}{k_b}/(2\frac{k_I}{k_b}-1)$. We note that we require $\abr{x^2}\geq0$, i.e. that $2k_I/k_b > 1$. We also note that this condition is satisfied by all empirical BMI distributions in the NU, NHANES and BRFSS data sets. Similarly, multiplying \eq{peqSolve2} by $x^2$, re-arranging, integrating, and solving for $\abr{x^3}$,  yields $\abr{x^3} = \frac{k_I}{k_b}/(\frac{k_I}{k_b} - 1) \abr{x^2}\xsp$.

The mean, variance, skewness, and mode skewness can now be computed using the following relations to the mode and the first three moments.
\begin{align*}
	\mbox{mean} &= \abr{x}\\
	\mbox{variance} &= \abr{x^2} - \abr{x}^2\\
	\mbox{skewness} &= \frac{\abr{x^3} - 3\abr{x}(\mbox{variance}) - \abr{x}^3 }{(\mbox{variance})^\frac{3}{2}}\mbox{, and}\\
	\mbox{mode skewness} &= \frac{\mbox{mean}-\mbox{mode}}{(\mbox{variance})^{\frac{1}{2}}} \ .
\end{align*}

\subsection{Solving \eq{peqSoln} for $p_{eq}(x)$ (social model, $k_S\ne0$)}
\label{A:peqIter}

In the case of the social model ($k_S\ne0$ in \eq{Mdetdyn}), \eq{peqSoln} does not provide a closed-form solution for the equilibrium distribution.
However, the stationary solution to \eq{2peqSolve} is given implicitly by \eq{peqSoln}, i.e., \eq{peqSoln} becomes
\[
	p_{eq}(x) \propto p_{eq}^{(0)}(x;k_I/k_b, \xsp) \exp\left( 2\frac{k_S}{k_b} \int_0^x \int_0^\infty \frac{\phi_{\tilde{x},\sigma}(\hat{x})(\hat{x} - \tilde{x})\left(1 - \half\frac{(\hat{x} - \tilde{x})^2}{\sigma^2} \right) p_{eq}(\hat{x})}{\tilde{x}^2} d\hat{x} d\tilde{x} \right) \ ,
\]
where we consider the continuum limit and the discrete sum in \eq{Mdetdyn} has been replaced by an integral over the population with distribution $p_{eq}(x)$.
In order to solve for $p_{eq}(x)$ numerically, we implement the following iterative scheme in which we discretize the iterative approximations $p_{eq}^{(i)}(x)$ and approximate the double integral numerically.
\begin{align}
	\nonumber
	p_{eq}^{(n+1)}(x) &= p_{eq}^{(n+1)}(x; k_I/k_b, \xsp, k_S/k_b, \sigma) \\
	\label{eq:peqIter}
    & \propto p_{eq}^{(0)}(x;k_I/k_b, \xsp) 
    \exp\left( 2\frac{k_S}{k_b} \int_0^x \int_0^\infty \frac{\phi_{\tilde{x},\sigma}(\hat{x})(\hat{x} - \tilde{x}) \left[1 - \half\frac{(\hat{x} - \tilde{x})^2}{\sigma^2} \right] p_{eq}^{(n)}(\hat{x})}{\tilde{x}^2} d\hat{x} d\tilde{x} \right)\ .
\end{align}
Let $m = 181$, $\Delta z = 0.5$, and and $\forall i = 1,2,\ldots,m: z_i = 10 + (i-1)\Delta z$. We set
\[
	p_{eq}^{(0)}(z_i) = \frac{z_i^{-2(k_I/k_b+1)}\exp\left(-2\frac{k_I}{k_b}\frac{\xsp}{z_i}\right)}{\sum_{j=1}^m z_j^{-2(k_I/k_b+1)}\exp\left(-2\frac{k_I}{k_b}\frac{\xsp}{z_j}\right)\Delta z} \ .
\]
We then set
\[
	p_{eq}^{(n+1)}(z_i) = p_{eq}^{(0)}(z_i) \exp\left(2\frac{k_S}{k_b} \sum_{k=1}^i\sum_{j=1}^m \half\left(1 + \mathbb{I}_{\{k<i\}}\right)\frac{\phi_{z_k,\sigma}(z_j)(z_j-z_k)\left(1-\half\frac{(z_j-z_k)^2}{\sigma^2}\right) p_{eq}^{(n)}(z_j)}{z_k^2}\Delta z^2 \right)
\]
where $\mathbb{I}_{\{X\}} = 1$ if $X$ is true and $\mathbb{I}_{\{X\}} = 0$ otherwise, and where we terminate the iterative process once
\[
	\left\|\frac{p_{eq}^{(n+1)} - p_{eq}^{(n)}}{m}\right\|_2 = \sqrt{ \sum_{i=1}^m \left( \frac{p_{eq}^{(n+1)}(z_i)-p_{eq}^{(n)}(z_i)}{m} \right)^2} < 10^{-12}\ .
\]

\subsection{Fitting distribution functions to empirical BMI distributions}
\label{A:dist_fit}

Suppose that $f(x;\theta)$ is a probability density function with parameters $\theta$. We fit $f(x;\theta)$ to empirical BMI data measurements $\{x_i\}_{i=1}^N$ using the principle of maximum likelihood parameter estimation. Specifically, we set
$$
	\hat{\theta} = \argmax_{\theta}  \prod_{i=1}^N f(x_i;\theta),
$$
where $\mathcal{L}_f(\theta|x) = \prod_{i=1}^N f(x_i;\theta)$ is called the likelihood function. In Matlab we perform this optimization using the Matlab function \emph{fminsearch} to solve the equivalent optimization problem 
$$
	\hat{\theta} = \argmin_{\theta} -\log(\mathcal{L}_f(\theta|x)) = \argmin_{\theta} - \sum_{i=1}^N \log\left[ f(x_i|\theta) \right]
$$

We note that we compute a separate set of parameters for each year of BMI data.

\subsection{Estimating parameters $(\hat{k}_I,\xsphat, \hat{k}_S, \hat{\sigma})$ in $a(x)$ (\eq{Mdetdyn}) and $\widehat{\sqrt{k_b}}$ in $b(x)$ (\eq{Mb})}
\label{A:ind_params}

In this section we describe how we estimate the parameters for the drift and diffusion curves (solid black) in Fig.~\ref{fig:2011_Ya_Yb}, fitted to the individual-level NU and NHANES data for all available years. The fitted parameters are presented in Extended Data Table \ref{tab:fitaMain}.

Consider individual $i$ from survey year $t=t_1$ with BMI measurements at times $t_1$ and $t_2 = t_1+\Delta t$, i.e. with BMI measurements $x_i(t_1)$ and $x_i(t_2)$. We denote the change in BMI by $\Delta x_i(t_t) = x_i(t_2)-x_i(t_1)$. For $\epsilon > 0$ we define
\begin{subequations}
\begin{align}
	\hat{a}(x_i(t);\epsilon) &= 
	\frac{\sum_{j:|x_j(t)-x_i(t)|<\epsilon} \frac{\Delta x_j(t)}{\Delta t} }{N(x_i(t), \epsilon)},\ \mbox{and} \label{eq:aapprox}\\
	\hat{b}(x_i(t);\epsilon) &= 
	\sqrt{ \frac{\sum_{j:|x_j(t) -x_i(t)|<\epsilon} \frac{\Delta x_j^2(t)}{\Delta t} }{N(x_i(t),\epsilon)} - \Delta t\left[\frac{\sum_{j:|x_j(t)-x_i(t)|<\epsilon} \frac{\Delta x_j(t)}{\Delta t} }{N(x_i(t), \epsilon)}\right]^2}
	\label{eq:bapprox}\ ,
\end{align}
\end{subequations}
respectively, where $N(x_i(t),\epsilon)$ is the number of individuals $j$ with $|x_j(t)-x_i(t)|<\epsilon$, i.e., 
$$
	N(x_i(t),\epsilon) = \sum_{j:|x_j(t) - x_i(t)| < \epsilon}1\ .
$$
We note that in order to reduce computation time we do not compute \eqs{aapprox}{bapprox} for each individual $i$ separately. Instead, we compute \eqs{aapprox}{bapprox} on the grid $\{10,10.1,10.2,\ldots,100\}$ and then evaluate $\hat{a}(x_i(t);\epsilon)$ and $\hat{b}(x_i(t);\epsilon)$ using linear interpolation.

To estimate $\sqrt{k_b}$ we compute $\hat{b}(x_i(t);\epsilon)$ from BMI data and regress it on $x_i(t)$. To estimate the remaining parameters we define the objective function
\[
	\mathcal{S}(k_I,\xsp, k_S, \sigma;\epsilon) = \sqrt{\frac{\sum_{i,t} \left[ \hat{a}(x_i(t);\epsilon) - a(x_i(t)) \right]^2}{\sum_{i,t} 1}}
\]
and solve the optimization problem
\begin{equation}
	\label{eq:params}
	\left( \hat{k}_I, \xsphat, \hat{k}_S, \hat{\sigma} \right) = \argmin_{(k_I,\xsp,k_S,\sigma)}\ \mathcal{S}(k_I,\xsp, k_S, \sigma;\epsilon),
\end{equation}
where we have suppressed the dependence of $(\hat{k}_I, \xsphat, \hat{k}_S, \hat{\sigma})$ on $\epsilon$ for convenience of notation. Recall from \eq{detdyn} that 
$$
	a(x_i(t)) = k_I (\xsp-x_i(t)) + k_S\frac{du_S}{dx_i}(x_i(t)),
$$
where 
\begin{align*}
	\frac{du_S}{dx_i}(x_i(t)) = \frac{1}{N(t)}\sum_{j=1}^{N(t)} \phi_{x_i(t),\sigma}(x_j(t))(x_j(t) - x_i(t))\left(1-\half\frac{(x_j(t)-x_i(t))^2}{\sigma^2}\right),
\end{align*}
and where $N(t)$ is the number of observations in survey year $t$. 
Observe that, for fixed $\sigma$, the objective function $\mathcal{S}(k_I,\xsp, k_S, \sigma;\epsilon)$ is the objective function for the linear regression of $\hat{a}(x_i(t);\epsilon)$ on $-x_i(t)$, $\frac{du_S}{dx_i}(x_i(t);\sigma)$, and a constant. It follows, therefore, that there is a unique $(\tilde{k}_I(\sigma), \tilde{x}^\star(\sigma), \tilde{k}_S(\sigma))$ that solves
\begin{equation}
	\label{eq:params_no_sigma}
	\left(\tilde{k}_I(\sigma), \tilde{x}^\star(\sigma), \tilde{k}_S(\sigma)\right) = \argmin_{(k_I,\xsp,k_S)} \mathcal{S}(k_I,\xsp, k_S, \sigma;\epsilon)\ ,
\end{equation}
and that can be computed using linear regression. Solving the optimiation problem in \eq{params} is now reduced to a one dimensional problem, i.e., we solve
\[
	\hat{\sigma} = \argmin_{\sigma}\ \mathcal{S}\left(\tilde{k}_I(\sigma), \tilde{x}^\star(\sigma), \tilde{k}_S(\sigma), \sigma\right)
\]
and set $(\hat{k}_I, \xsphat, \hat{k}_S) = (\tilde{k}_I(\hat{\sigma}), \tilde{x}^\star(\hat{\sigma}), \tilde{k}_S(\hat{\sigma}))$.

We estimate the parameters $k_I$ and $\xsp$ with $k_S=0$ by regressing $\hat{a}(x_i(t);\epsilon)$ on $-x_i(t)$ and a constant. Note that when $k_S=0$ the parameter $\sigma$ is undetermined.

We note that the methodology presented in this section can only be applied to NHANES and NU BMI data, because these are the only data sets that have information on how individuals' BMI changes over time. We are able to compute $\Delta x_i(t)$ for individuals $i$ in the NHANES data set with self-reported weights WHD010 (current, i.e. at time $t_2$) and WHD050 (one year prior to survey, i.e. at time $t_1$), and with self-reported height WHD020. For convenience we set $t=1999$ for the 1999-2000 NHANES survey, $t=2001$ for the 2001-2002 NHANES survey, etc... We note that for the NHANES data $\Delta t =  1$. For NU data we also consider $\Delta t = 1$, i.e., we consider individuals for whom we can compute BMI in two consecutive years. NU data for individuls in consecutive years exists for years 
\[
	t \in \{ 1996, \ldots, 2013\}.
\]
As above, if multiple weight measurements are present in year $t$ we calculate the BMI for that year using the average weight in year $t$, whereas if multiple heights measurements are present then we calculate BMI using the average height (where the average is taken over all years). We note that for both data sets we use $\epsilon=\frac{1}{2}$ to compute \eqs{aapprox}{bapprox}.

All computations are performed in Matlab. Regressions are performed using the Matlab function \emph{regress}. Optimization are performed using the Matlab function \emph{fminsearch}.
%

\subsection{Akaike Information Criterion}
\label{sec:AIC-app}

We give a brief overview of maximum likelihood estimation and the Akaike Information Criterion.

\subsubsection{Maximum Likelihood Estimation}

Suppose that we have independently and identically distributed (IID) data $\{x_i\}_{i=1}^N$ that are drawn from the unknown probability distribution $p(x)$. Suppose also that we are attempting to model the unknown probability distribution $p(x)$ by the family of parametric probability distribution functions $\{f(x|\theta)\}_{\theta}$, i.e. our goal is to find the $\hat{\theta}$ such that, of all the functions in $\{f(x|\theta)\}_{\theta}$, $f(x|\hat{\theta})$ is the ``best''  approximation to $p(x)$. The maximum likelihood estiamtor (MLE) $\hat\theta$ is the parameter that maximizes the likelihood function $\mathcal{L}_f(\theta|x)$, i.e. 
\begin{equation} 
	\label{eq:MLE}
	\hat\theta = \argmax_{\theta} \underbrace{\prod_{i=1}^N f(x_i|\theta)}_{=\mathcal{L}_f(\theta|x)}.
\end{equation}

We note that the relative likelihood function 
$$
	r = \frac{\mathcal{L}_f(\theta|x)}{\mathcal{L}_f(\hat\theta|x)}
$$
is interpreted as follows: $f(x|\theta)$ is $r$ times as likely as $f(x|\hat\theta)$ to be the ``best'' approximation to $p(x)$. In this case, ``best'' is in the context of maximizing the likelihood function.

\subsubsection{Akaike Information Criterion}

The Akaike Information Criterion (AIC) is a generalization of the principle of maximum likelihood estimation. An equivalent formulation of the MLE given in Eq.~\eqref{eq:MLE} is given by maximizing the average log-likelihood function $\mathcal{S}_N(f(\cdot|\theta))$, i.e. 
$$
	\hat\theta = \argmax_{\theta} \frac{1}{N} \log\left(\mathcal{L}_f(\theta|x)\right) =  \argmax_{\theta} \underbrace{ \frac{1}{N}  \sum_{i=1}^N \log( f(x_i|\theta) ) }_{=\mathcal{S}_N(f(\cdot|\theta))}.
$$

It can be shown that the mean log-likelihood function $\mathcal{S}_N(f(\cdot|\theta))$ converges with probability 1 to 
$$
	\mathcal{S}(p;f(\cdot|\theta)) = \int p(x) \log(f(x|\theta)) dx.
$$
From this quantity we define the Kullback-Leibler mean information for the discrimination between $p(x)$ and $f(\cdot|\theta)$
$$
	I(p;f(\cdot|\theta)) = \mathcal{S}(p;p) - \mathcal{S}(p;f(\cdot|\theta)),
$$
which can be shown to be non-negative, with $I(p;f(\cdot|\theta)) = 0 \iff f(x|\theta) = p(x)$ almost everywhere. Roughly speaking, $I(p;f(\cdot|\theta))$ can be interpreted as the amount of information lost when $f(\cdot|\theta)$ is used to approximate $p(x)$. This quantity induces a natural model selection criterion, i.e. we select the model that minimizes $I(p;f(\cdot|\theta))$.

{\bf Remarks:}
\begin{enumerate}
	\item $\mathcal{S}(p;f(\cdot|\theta))$ can be approximated by $\mathcal{S}_N(f(\cdot|\theta))$, which can be computed from the data without knowing the ``true'' distribution $p(x)$.
	\item Setting $\hat\theta = \argmin_{\theta} I(p;f(\cdot|\theta))$, equivalently $\hat\theta = \argmax_{\theta} \mathcal{S}(p;f(\cdot|\theta)) \approx \argmax_\theta \mathcal{S}_N(f(\cdot|\theta))$ recovers the MLE.
\end{enumerate}	

The key observation for the establishment of the AIC criterion is that the quantity $I(p;f(\cdot|\theta))$ can be approximated as follows. Suppose that the true model is $p(x) = f(x|\theta_0)$ for some $\theta_0\in\Theta$ and suppose that $\hat\theta$ is the MLE for the model restricted to some $k$-dimensional subspace $\Theta'\subset\Theta$, i.e.
$$
	\hat\theta = \argmax_{\theta\in\Theta'\subset\Theta} \mathcal{L}_f(\theta|x).
$$ 
Then it can be shown that
$$
	\mathbb{E}\left[2N\ I(p;f(\cdot|\hat\theta))\right] = \mathbb{E}\left[2N\ I(f(\cdot|\theta_0);f(\cdot|\hat\theta))\right] = c + 2k - 2\sum_{i=1}^N \log\left(f(x_i|\hat\theta)\right) = c + \underbrace{2k - 2\log\left( \mathcal{L}_f(\hat\theta|x) \right)}_{=AIC(f(\cdot|\hat\theta))},
$$
where $c$ is a constant, $k$ is the dimension of $\Theta'$ (i.e., the number of parameters in the model), and where $AIC(f(\cdot|\hat\theta)) = 2k - 2\log \left( \mathcal{L}_f(\hat\theta|x) \right)$ is the Akaike Information Criterion (AIC). It follows that minimizing $I(p;f(\cdot|\theta))$ is equivalent to minimizing the AIC.
A key point is that $k$ is the number of parameters in the model, and the AIC deals with the trade-off between the goodness of fit of the model and the complexity (number of parameters) of the model.

We now would like to generalize the likelihood ratio introduced above. Suppose that we compute the AIC for two different models resulting in AIC values $AIC_1 = 2k_1 - 2\log\left( \mathcal{L}_1 \right)$ and $AIC_2 = 2k_2 - 2\log\left(\mathcal{L}_2\right)$ with $AIC_1 < AIC_2$. Then the relative likelihood ratio
$$
	r = \exp\left( \frac{AIC_1 - AIC_2}{2} \right) = \exp\left(k_1 - k_2\right) \frac{\mathcal{L}_2}{\mathcal{L}_1} 
$$
is interpreted as follows: model 1 is $r$ times as likely to be the ``best'' approximation to the true distribution than model 2. In this case, ``best'' is in the context of minimizing the AIC (i.e. minimizing the loss of information when using models 1/2 to approximate the ``true'' distribution $p(x)$).

\newpage
\section{Data and Code Files Made Available with this Manuscript} \label{A:files}

The following files can be downloaded from H. De Sterck's website via \url{http://tinyurl.com/BMI-code-data}.

\subsection{Matlab Code}

The results presented in this paper were generated using the following four Matlab m-files.
\begin{enumerate}
	\item \emph{BMI\_Master.m}: Executes files \emph{fitBMIdistn.m} and \emph{fitAB.m}, see below.
	\item \emph{fitBMIdistn.m}: Performs population-level calculations, i.e. fits nonsocial model $p_{eq}^{(0)}(x)$, social model $p_{eq}(x)$, log-normal $\flog$, and skew normal $\fskew$ distributions to empirical BMI distribution data. 
	\item \emph{fitAB.m}: Performs individual-level calculations, i.e. computes coefficients a(x) and b(x) from year-over-year change in BMI data.
\end{enumerate}

We note that both \emph{fitBMIdistn.m} and \emph{fitAB.m} make use of the m-file \emph{save2pdf.m} \citeSI{GabeHoffman} when saving figures.

\subsection{NU, NHANES, and BRFSS BMI Data Files}

\begin{enumerate}
	\item NU data are stored in \emph{NU.csv} comma separated values (CSV) format. This file contains three columns: year $t$, BMI in year $t$, BMI in year $t+1$.
		\begin{itemize}
			\item \emph{Note}: When BMI in year $t+1$ is unavailable then the entry in the third column is -1. 
		\end{itemize}
	\item NHANES data are either self-reported (used to calculate year-over-year change in BMI) or directly measured (used to compute BMI distributions) data. We associate NHANES data from survey 1999-2000 with $t=1999$, from survey 2001-2002 with $t=2001$, and so on.
		\begin{enumerate}
			\item Self-reported NHANES data are stored in \emph{NHANES\_SR.csv} in CSV format. This file contains three colums: year $t$, BMI in year $t-1$ (i.e. year prior to the interview), and BMI in year $t$ (i.e. at time of interview).
				\begin{itemize}
					\item \emph{Note}: Because self-reported NHANES data are only used for individual-level computations, \emph{NHANES\_SR.csv} only records data from respondents who reported both (a) BMI at time of NHANES interview and (b) BMI one year prior to NHANES interview.
				\end{itemize} 
			\item Directly measured NHANES data are stored in \emph{NHANES\_DM.csv} in CSV format. This file contains two columns: year $t$ and BMI in year $t$.
		\end{enumerate}
	\item BRFSS data are stored in \emph{BRFSS\_BMI.csv} in CSV format. This file contains two columns: year $t$, BMI in year $t$.
\end{enumerate}

\bibliographystyleSI{naturemag}
\bibliographySI{biblio3}

\end{document}